\documentclass[sigconf]{acmart}

\usepackage[utf8]{inputenc}
\usepackage{booktabs}
\usepackage{graphicx}
\usepackage{subfigure}
\usepackage{amsmath}
\usepackage{booktabs}
\usepackage{multirow}
\usepackage{multicol}
\usepackage{xcolor}
\usepackage[linesnumbered,ruled]{algorithm2e}

\SetCommentSty{mycommfont}
\DeclareMathOperator*{\argmax}{arg\max}
\newcommand{\RNum}[1]{\uppercase\expandafter{\romannumeral #1\relax}}

\copyrightyear{2019}
\acmYear{2019}
\setcopyright{iw3c2w3}
\acmConference[WWW '19]{Proceedings of the 2019 World Wide Web Conference}{May 13--17, 2019}{San Francisco, CA, USA}
\acmBooktitle{Proceedings of the 2019 World Wide Web Conference (WWW'19), May 13--17, 2019, San Francisco, CA, USA}
\acmDOI{10.1145/3308558.3313455}
\acmISBN{978-1-4503-6674-8/19/05}

\begin{document}

\title[Aggregating E-commerce SRs from Heterogeneous Sources via HRL]{Aggregating E-commerce Search Results from Heterogeneous Sources via Hierarchical Reinforcement Learning}

\author{Ryuichi Takanobu}
\authornote{Both authors contributed equally to this research.}
\affiliation{
  \institution{Institute for AI,
  State Key Lab of Intelligent Technology \& Systems,
  DCST, Tsinghua University}
  \city{Beijing}
  \country{China}
}
\email{gxly15@mails.tsinghua.edu.cn}

\author{Tao Zhuang}
\authornotemark[1]
\affiliation{
  \institution{Alibaba Group}
  \city{Hangzhou}
  \country{China}
}
\email{zhuangtao.zt@alibaba-inc.com}

\author{Minlie Huang}
\authornote{Corresponding author: Minlie Huang.}
\affiliation{
  \institution{Institute for AI,
  State Key Lab of Intelligent Technology \& Systems,
  DCST, Tsinghua University}
  \city{Beijing}
  \country{China}
}
\email{aihuang@tsinghua.edu.cn}

\author{Jun Feng}
\authornote{Participated in this work while at Tsinghua University.}
\affiliation{
  \institution{State Grid Zhejiang Electric Power Co., LTD}
  \city{Hangzhou}
  \country{China}
}

\author{Haihong Tang}
\affiliation{
  \institution{Alibaba Group}
  \city{Hangzhou}
  \country{China}
}

\author{Bo Zheng}
\affiliation{
  \institution{Alibaba Group}
  \city{Hangzhou}
  \country{China}
}

\renewcommand{\shortauthors}{Ryuichi Takanobu, et al.}

\begin{abstract}
In this paper, we investigate the task of aggregating search results from heterogeneous sources in an E-commerce environment.
First, unlike traditional aggregated web search that merely presents multi-sourced results in the first page, this new task may present aggregated results in all pages and has to dynamically decide which source should be presented in the current page. Second, as pointed out by many existing studies, it is not trivial to rank items from heterogeneous sources because the relevance scores from different source systems are not directly comparable.
To address these two issues, we decompose the task into two subtasks in a hierarchical structure: a high-level task for {\it source selection} where we model the sequential patterns of user behaviors onto aggregated results in different pages so as to understand user intents and select the relevant sources properly; and a low-level task for {\it item presentation} where we formulate a slot filling process to sequentially present the items instead of giving each item a relevance score when deciding the presentation order of heterogeneous items.
Since both subtasks can be naturally formulated as sequential decision problems and learn from the future user feedback on search results, we build our model with hierarchical reinforcement learning. Extensive experiments demonstrate that our model obtains remarkable improvements in search performance metrics, and achieves a higher user satisfaction.
\end{abstract}



\maketitle

\section{Introduction}
The process of aggregating search results from heterogeneous sources is usually referred to as \textit{aggregated search} \cite{kopliku2014aggregated}. Different from a meta-search engine or search federation which assumes that all distributed sources contain homogeneous content and applies the same ranking function to each source for result fusion \cite{arguello2017aggregated}, an aggregated search system is expected to support \textbf{heterogeneous information seeking} from different sources. When a user issues a query, the search system should integrate relevant results from each specialized search system, called \textit{vertical}, into one Search Engine Result Page (SERP).

\begin{figure}[!htb]
\centering
\includegraphics[width=.9\linewidth]{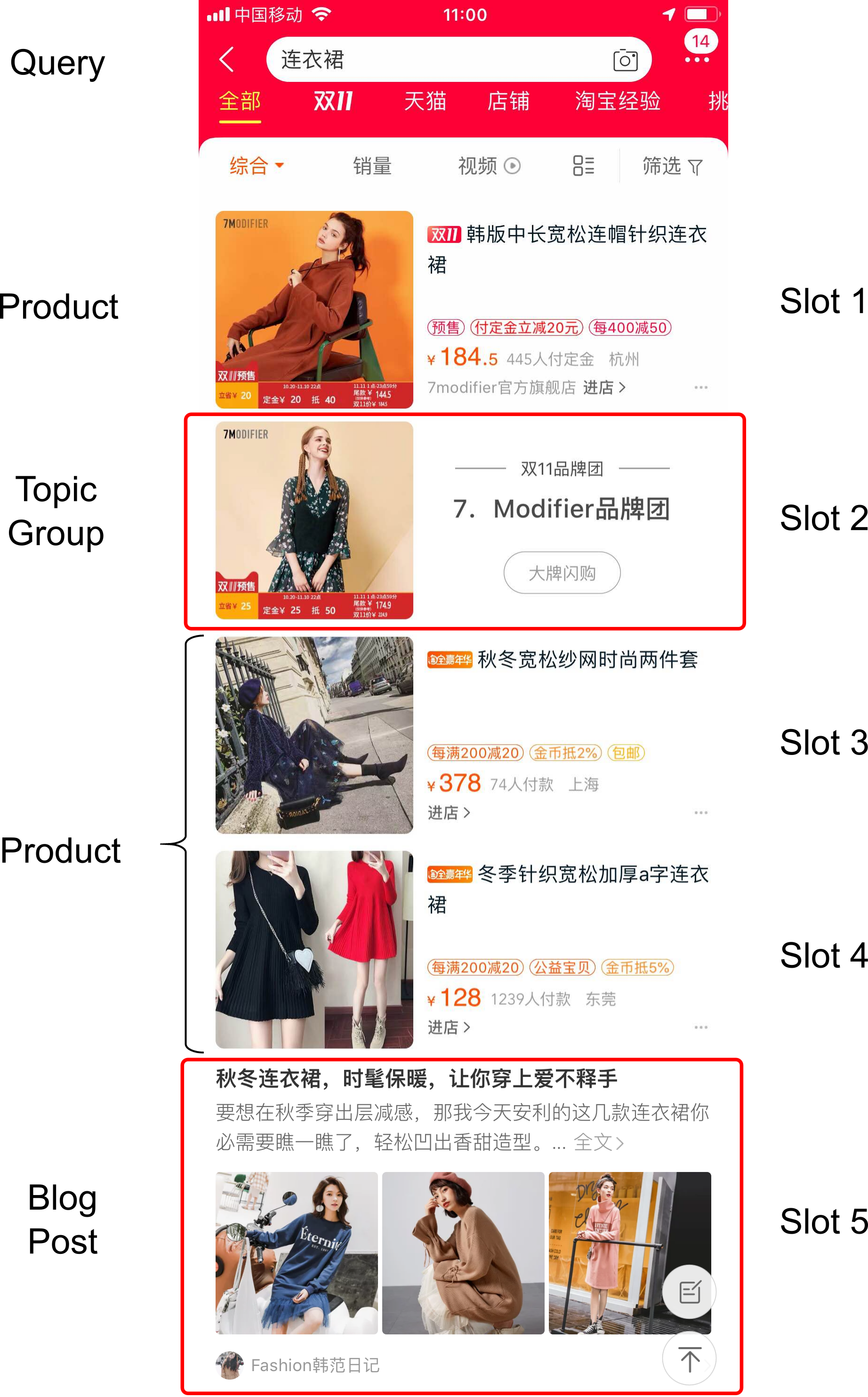}
\caption{An aggregated search example for the query ``dress'', where the topic group from the \textit{topic} vertical is shown in the 2nd position, and the blog post from the \textit{blog} vertical in the 5th position.}
\label{vertical}
\end{figure}

Existing research on search result aggregation has focused on web search \cite{arguello2009sources,diaz2009integration,long2014relevance}, while in this paper, we study the task in E-commerce search with one of the largest E-commerce search services in the world: \textit{Taobao.com}. Fig. \ref{vertical} illustrates that traditional E-commerce search can be augmented with \textit{blog posts} that share purchasing experiences, or \textit{topic groups} that cluster products with the same brand or from the same shop to facilitate finding similar products.
As a matter of fact, user study in Taobao Search shows that mixing products with topic groups and blog posts in a SERP can greatly improve the user experiences of product search and online shopping. In this study, we aggregate two other verticals in product search, namely the topic and blog verticals.
Given a query, these verticals return topic groups and blog posts respectively.

In web search \cite{arguello2009sources,diaz2009integration,long2014relevance},  result aggregation is performed only once at the first page for each query.
While in E-commerce search, aggregation is performed multiple times for a query, one for each page.
The aim of page-wise aggregation in E-commerce search is to keep diversity to attract and retain diverse customers. This is a major difference from previous web search aggregation in which a single decision is made only for the first page.

In this paper, we decompose search result aggregation in E-commerce search into two subtasks: \textit{source selection} and \textit{item presentation}.
\textit{Source selection} decides whether to present the search results of a certain source type in the current page, which depends on the user behaviors onto the items that have been already presented in previous pages.
For example, if a user issued the query “sport shoes”, and clicked several products with the same brand, presenting a topic group with the same brand in the next page may be a good decision.
Then the validity of search results can be examined afterwards by analyzing the long-term gains like click through rate (CTR). So this process is a sequence of decisions and can be modeled with reinforcement learning (RL) \cite{sutton1998reinforcement}, which helps better understand user intents to present more relevant aggregated search results.

The other subtask, \textit{item presentation}, is to decide the presentation order of the items from heterogeneous sources, which requires to estimate the relevance scores of all items in a unified way. However, items from different sources have different relevance estimation models \cite{arguello2011learning,ponnuswami2011composition,jie2013unified}, thereby making the scores not comparable between different sources. This is termed the \textit{relevance ranking} issue.
To avoid this, we formulate the item presentation task as a \textit{slot filling} problem where the training signal comes directly from the user feedback.
In this formulation, each display position in a page is regarded as a slot, and the slots in a page will be filled sequentially.
At each position, an \textit{item presenter} is trained to choose the most relevant item from the candidate sources, and then fills in the slot with the selected item.
Once again, the slot filling process is consistent with the sequential nature of user behaviors, which can be viewed as a sequential decision problem and handled by RL.

Given the above decomposition, we propose a hierarchical reinforcement learning (\textbf{HRL}) model \cite{barto2003recent} consisting of two components: a source selector that decides which sources should be selected at the current page, and an item presenter that decides the presentation order of the items selected from the selected sources in a page.
The model fully utilizes the sequential characteristics of user behaviors across different pages to decide both the sources and items to be presented. The hierarchical framework enables the model to utilize the immediate user feedback in the current page as well as the future user feedback on the entire search session.

To summarize, our main contributions are as follows:

\begin{itemize}
    \item We propose a novel search result aggregation method that formulates a semi-Markov decision process which is composed of a high-level policy for \textit{source selection} and a low-level policy for \textit{item presentation}.
    \item We present aggregated search results for each page during source selection and capture the sequential patterns of user behaviors onto aggregated items in different pages.
    \item We formulate the item presentation subtask as a slot filling problem in order to avoid the relevance ranking issue on displaying heterogeneous items.
\end{itemize}

Results demonstrate that our proposed model can improve user experience and make significant advancement on marketing metrics like CTR and GMV over other baselines.

\section{Preliminaries}\label{sec:preliminary}
\subsection{Reinforcement Learning}

Reinforcement learning (RL) \cite{sutton1998reinforcement} is a learning paradigm that an agent learns from the interactions with the environment through sequential exploration and exploitation. Generally, RL follows a Markov Decision Process (MDP) formulation.

Let an MDP be defined as $< \mathcal{S}, \mathcal{A}, \mathcal{P}, r, \gamma >$, where $\mathcal{S}$ denotes the state space, $\mathcal{A}$ denotes the set of possible actions, $\mathcal{P} : \mathcal{S} \times \mathcal{A} \times \mathcal{S} \rightarrow [0, 1]$ denotes the transition function to generate the next state from current state-action pair, $r : \mathcal{S} \times \mathcal{A} \times \mathcal{S} \rightarrow \mathbb{R}$ denotes the reward function and $\gamma \in [0, 1]$ is an discount factor . Given an MDP, a trajectory $\tau = s_0, a_0, r_0, s_1, \cdots$ that starts from state $s_0$ is sampled from a policy $\pi : \mathcal{S} \times \mathcal{A} \rightarrow [0, 1]$ that specifies the action. The aim of RL is to find an optimal policy $\pi^*$ which maximizes the expected cumulative reward (return) $R_t = \mathbb{E}[\sum_{k=0}^\infty \gamma^k r_{t+k}]$ at each time step $t$. For an agent following a policy $\pi$, the Q-value of the state-action pair $(s, a)$ is defined as $Q_\pi (s,a) = \mathbb{E} [R_t | s_t = s, a_t = a, \pi] = \mathbb{E}_{s' \sim \mathcal{P}(\cdot |s, a)} [r + \gamma \mathbb{E}_{a' \sim \pi(s')}[Q_\pi (s',a')] | s, a]$, which measures the average discounted long-term rewards.
In the practice of training, the action is often selected by an $\epsilon-greedy$ policy that follows the greedy policy $a = \argmax_{a'} Q(s, a')$ with probability $1 - \epsilon$ for exploitation and selects a random action with probability $\epsilon$ for exploration, in order to collect desired transitions for learning.

With the advance of deep neural networks, deep reinforcement learning (DRL) has been widely used in various search tasks in recent two years \cite{xia2017adapting,oosterhuis2018ranking,feng2018learning}.

\subsection{Options Framework}
\textit{Options framework} \cite{sutton1999between} extends the usual MDP so that it involves temporal abstractions over the action space in the context of HRL. At each step, the HRL agent chooses either a ``one-step'' action (primitive action) or a ``multi-step'' action (option). An option $o = (\pi, \mathcal{I}, \beta) \in \mathcal{O}$ defines a policy $\pi$ over actions (either primitive or option), includes an initiation set $\mathcal{I} \subset \mathcal{S}$ that option $o$ is available \textit{iff} state $s \in \mathcal{I}$ , and can be terminated according to a stochastic function $\beta : \mathcal{S} \rightarrow [0, 1]$. An MDP endowed with a set of options is extended to a Semi-Markov Decision Process (SMDP). Given an SMDP, a trajectory $\tau$ is sampled from a policy $\mu$ over options. Each option $o_k$ launches a new subtask where a sub-trajectory $\tau_k$ is sampled from the corresponding policy $\pi$ over actions. The agent cannot select next option (possibly in a recursive way) until current option terminates according to the termination function $\beta$.

Recently, several studies have demonstrated that combining DRL with predefined subgoals delivers promising results in challenging environments like Atari \cite{kulkarni2016hierarchical}, Minecraft \cite{tessler2017deep}, relation extraction \cite{takanobu2019hierarchical} and task-oriented dialogues \cite{peng2017composite}.

\section{Related Work}\label{sec:related_work}

\begin{figure}[!htp]
    \centering
    \includegraphics[width=.9\linewidth]{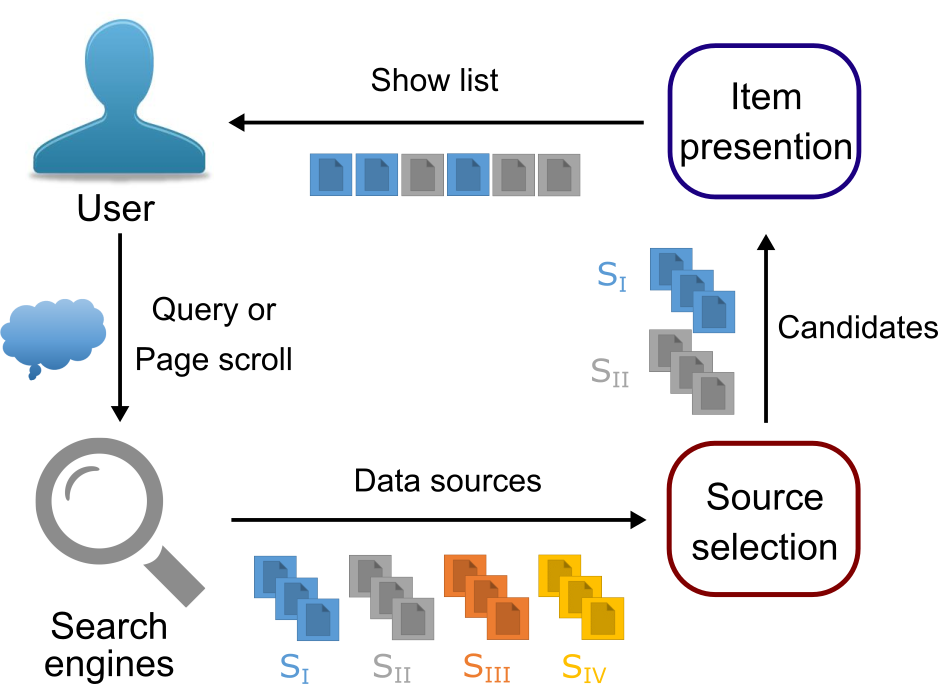}
    \caption{Task flow of search aggregation on heterogeneous sources.}
    \label{flow}
\end{figure}

In recent years, search engines are able to search heterogeneous sources that contain different types of contents (e.g. news, maps, videos, images). Studies have shown that the aggregated view helps expose the content in other sources, thus offering the chance for the user to explore different sources in search\cite{chuklin2013evaluating,bron2013aggregated}. Existing aggregated search systems, which merge all heterogeneous results from various search services in one SERP, mostly adopt a pipeline architecture as illustrated in Fig. \ref{flow}. In general, the architecture contains two key components, one is source selection which decides the vertical sources from which items should be ranked, and the other is item presentation which decides the presentation order of the heterogeneous items.

\subsection{Vertical Selection}

The first subtask of aggregated search is vertical selection, also called source selection. The goal of vertical selection is to decide which verticals are to be presented in a SERP, e.g. whether it is necessary to present the videos of ``violin'' when a user types query ``violin''. Several works \cite{zhou2013vertical,liu2015influence,turpin2016blend} have found that participants rated the system negatively when irrelevant vertical results were presented in an aggregated SERP. It is thus essential to evaluate the relevance of a source to a query.

Most existing approaches use a binary classifier for each vertical to make a decision on whether to present a vertical result in a page \cite{hong2010joint,tsur2016identifying,levi2018selective}. From this perspective, each classifier can adopt a different feature representation and focus on the features that are uniquely predictive for its corresponding vertical. Human judges are employed to generate ground-truth labels for the verticals in some approaches \cite{arguello2009sources,diaz2009integration}. Other approaches \cite{ponnuswami2011composition,jie2013unified} leverage user search logs and assign a binary label based on user behaviors.

As previous studies focused on web search, vertical selection is only launched one time when a user issues a query. For instance, the \textit{video} vertical will only appear once in the first SERP for query ``Titanic'', but no vertical will be shown again in the following pages unless the user types a new query.
However, the goal of E-commerce search is quite different from web search \cite{zhuang2018ranking}.
With respect to aggregated search, the main difference between web and E-commerce lies in that the latter needs to present the aggregated search results in each page to meet user search demand, which requires the system to capture the sequential patterns of user behaviors onto aggregated results.

\subsection{Vertical Presentation}

The second subtask, vertical presentation, or item presentation, is to rank the candidate items selected from verticals. For instance, should the videos of ``violin'' be positioned above or below the web links in a page? In general, the system presents the most relevant items at the top, which are more likely to be viewed by users \cite{sushmita2010factors,chen2015does}. Existing approaches to vertical presentation can be classified into three types: \textit{pointwise}, \textit{pairwise} and \textit{attention-based} approaches.

\textbf{Pointwise} approaches train an independent classifier per vertical to predict the relevance of the corresponding vertical.  \textit{Click models} were proposed to predict the user response (i.e. click or skip) to vertical results  \cite{wang2013incorporating,jie2013unified,wang2016beyond}. FCM \cite{chen2012beyond} considered the rank of the verticals, their visual salience, and the distance between them and the current item. \citeauthor{markov2014vertical} \shortcite{markov2014vertical} proposed a click model that estimated different probabilities for results above or below the vertical as well.
Others directly trained \textit{vertical-specific classifiers} to predict the relevance score of a source to a query, which was used to decide the position of each vertical \cite{si2003relevant,bota2014composite,ponnuswami2011model}. \citeauthor{ponnuswami2011composition} \shortcite{ponnuswami2011composition} trained a gradient boosted decision tree to learn relevance scores and ranked each vertical individually using a computed threshold. Although it is simple and intuitive to implement pointwise models, they suffer from the \textit{relevance ranking} issue that the prediction scores from individual classifiers are not directly comparable.

\textbf{Pairwise} approaches learn to predict the \textit{relative preference} between candidate sources to be displayed in a SERP. \citeauthor{arguello2011learning} \shortcite{arguello2011learning} trained one binary classifier for user preference per vertical pair, and derived a final ranking from the predicted preferences. The pairwise method solves the \textit{relevance ranking} issue, but in return, it requires a complex ranking principle among verticals and results in a large number of pairwise classifiers.

The recent study \cite{zhang2018relevance} adapted an \textbf{attention-based} method that applied different weights to different information sources for relevance estimation. It has gained a significant improvement on ranking relevance, however, it requires a large number of data annotations on relevance score to support model training.

\section{Hierarchical Search Aggregation}

\begin{figure}[!htp]
    \centering
    \includegraphics[width=\linewidth]{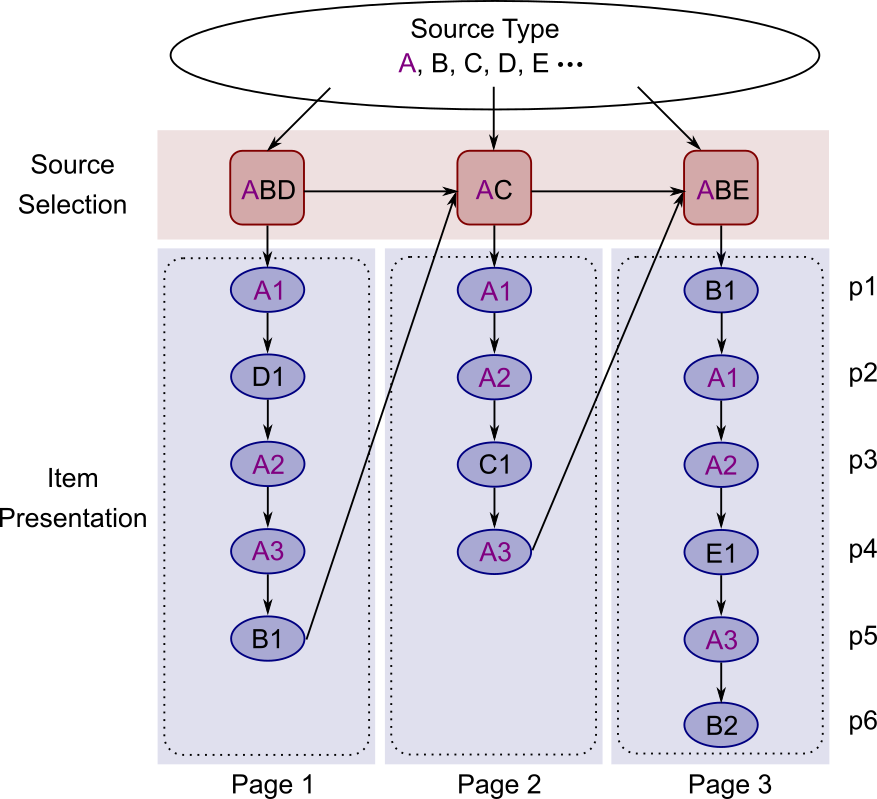}
    \caption{The hierarchical architecture on aggregated search. This example of user search session is composed of three search requests (page 1 to 3). $A$ stands for the products, and $B, C \cdots$ for the items of other source types.}
    \label{hrl}
\end{figure}

We develop a deep hierarchical reinforcement learning algorithm for aggregated search in this paper. The model consists of a high-level source selection policy and a low-level item presentation policy. The entire process works as shown in Fig. \ref{hrl}. When a new search request arrives, the source selector decides which sources to present on the current page. The item presenter then decides the presentation order of the relevant items from selected sources. The user search session continues when the user scrolls to the next page, and ends when he leaves the platform or types a new query. Both policies are trained with the DQN \cite{mnih2015human} algorithm. The rewards for policy training come from implicit user feedback, instead of any artificial metrics \cite{arguello2011methodology,zhou2012evaluating}, so that it can learn from user behaviors straightforwardly.

In the following subsections, we will present the details of the two policies, deep Q network architecture, and model optimization. The notations used in this paper are summarized in Table \ref{notation}.

\begin{table}[!htb]
    \centering
    \begin{tabular}{ll}
    \toprule
    Notation&  Description \\
    \midrule
    $x$                   &  Search request\\
    $\mathcal{SR}_j$      &  Set of all items of source type $j$ \\
    $m_j^k$               &  K-th item of source type $j$ \\
    $n_j(x)$              &  The number of items of type $j$ given the request $x$ \\
    $p$, $P$    &  Slot position on the SERP, Result presentation \\
    \midrule
    $s$, $\mathcal{S}$    &  State, State set \\
    $a$, $\mathcal{A}$    &  Primitive action, Primitive action set \\
    $o$, $\mathcal{O}$    &  Option, Option set \\
    $r^e$, $r^i$          &  Extrinsic reward, Intrinsic reward \\
    $\gamma$              &  Discount factor \\
    $R$                   &  Return (discounted cumulative reward) \\
    $\mu$, $\pi$          &  Source selector policy, Item presenter policy \\
    \bottomrule
    \end{tabular}
    \caption{Notations of aggregated search and HRL.}
\label{notation}
\end{table}

\subsection{Source Selection with High-level Policy}\label{subsec:high}

The high-level RL process aims to decide which sources should be selected in each page.
The high-level policy $\mu$ perceives the high-level state $s^e$ and selects an option $o$ that tells what sources to display. The option $o$ will trigger the low-level policy to decide how to present the items from these selected sources.
The high-level source selector tries to capture the sequential patterns of user behaviors onto aggregated results in different pages.

\subsubsection{Option}

An option $o \in \mathcal{O}$ refers to a multi-step action \cite{sutton1999between} that selects the relevant sources to display. The core search, which is the product search in our case, must be selected, while verticals can either be selected or not. So if there are $N$ verticals, the size of the option set is $|\mathcal{O}| = 2^N$. The agent samples an option $o$ according to $\mu$, then the control of the agent transfers to the low-level policy. The option lasts until the low-level RL process terminates, then the source selector continues to execute the next option.

\subsubsection{State}

The state $s_t^e \in \mathcal{S}$ of the high-level RL process at time step $t$ is represented by: 1) the search request $x$, 2) the triggered search source results $\mathcal{SR}(x)$, and 3) the \textit{latest} option $o_{t'}$ where $t - t'$ is the duration of $o_{t'}$.
Then the state vector $\phi(s_t^e)$ is the concatenation of the embedding vector of each part. It should be noted that the vector of source type $\mathcal{SR}_j$ is represented by the mean value of all item vectors of that type, i.e. $\textbf{e}(\mathcal{SR}_j(x)) = \frac{1}{n_j(x)} \sum_k \textbf{e}(m_j^k(x))$ where $\textbf{e}(m)$ is the item vector of $m$. In particular, the vector of a certain source is represented by zeros if the result of that type is empty. Then the vector of entire source results $\textbf{e}(\mathcal{SR})$ is derived from the concatenation of $\textbf{e}(\mathcal{SR}_j)$ from all source types.

\subsubsection{Extrinsic reward}

The extrinsic reward $r^e$ is given to the source selector to evaluate the global performance. It is derived from the mean value of the intrinsic return from the low-level process, which can be formulated as follows:
\begin{equation}
    r_t^e = \frac{1}{l} R_t^i = \frac{1}{l} \sum_{k=0}^{l-1} \gamma^k r_{t+k}^i,
\label{reward:high}
\end{equation}
where $l$ is the duration of the option $o_t$. The average return is calculated to prevent the tendency that the source selector greedily chooses all triggered sources to obtain more rewards. The definition of intrinsic reward will be introduced soon later.

\subsection{Item Presentation with Low-level Policy}

The low-level RL process aims to decide the presentation order of the items from the candidate sources $\overline{\mathcal{SR}}(x)$ chosen by the high-level source selector.
Note that the items from each source have been ranked according to its private vertical, and all items from verticals can be interleaved with results from other sources.
Each display position is regarded as a slot and the entire presentation subtask starts by filling an item into the top slot of the page, and then the second slot, etc.
At each step, the low-level RL policy $\pi$ takes the corresponding option $o$ and the low-level state $s^i$ as input, and outputs a primitive action $a$ that tells which source to display at the current position. As a result, the agent does not give a relevance score to each item thus avoiding the relevance ranking issue.

Considering that the source selector has notified which sources are selected to display, the number of slots on the page can be determined by $l = \sum_{\mathcal{SR}_j \in \overline{\mathcal{SR}}} n_j(x)$.
The agent reaches the subgoal when all the slots in a page have been filled, thus the number of slots is equal to the time duration of the option, and the termination function $\beta : \mathcal{S} \rightarrow \{0, 1\}$ is deterministic in our setting.
Once it is terminated, the control of the agent will return to the high-level process.

\subsubsection{Primitive action}

\begin{figure}[!htp]
    \centering
    \includegraphics[width=.85\linewidth]{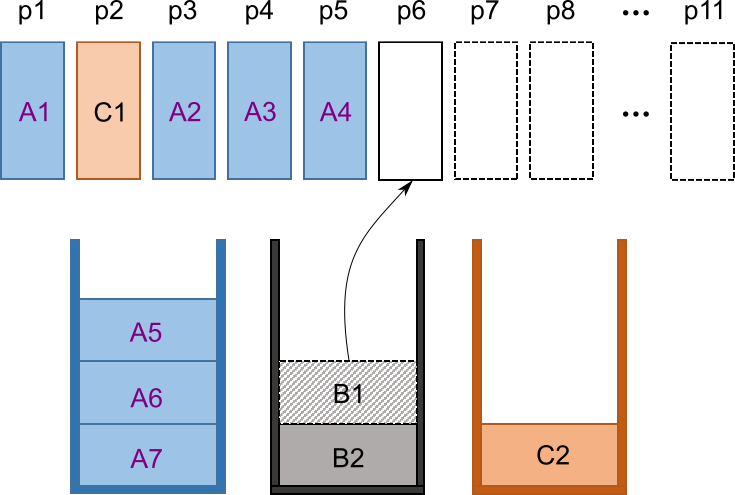}
    \caption{Illustration of the proposed slot filling scheme. There are three sources (A,B,C) and 7+2+2=11 slots in this example.}
    \label{slot}
\end{figure}

A primitive action $a \in \mathcal{A}$ refers to a one-step action that chooses a certain source type, so $|\mathcal{A}| = 1 + N$ given the core search and $N$ verticals.
Each source can be seen as a stack with its items in descending order of relevance to the search request $x$.
If a certain source is selected according to $\pi$, the top item at the stack of this source is displayed in the current position, as shown in Fig. \ref{slot}. When a source stack is empty, the source is excluded from the action space.
In this way, we simplify the relevance ranking problem into a slot filling process to mix different sources: we only need to decide from which source the top item should be selected, instead of evaluating the relevance of each source item.
Note that the relative order of the items within the same source type remains unchanged, the agent simply utilizes the existing ranking list.

\subsubsection{State}

The state $s_t^i \in \mathcal{S}$ of the low-level RL process at time step $t$ is represented by: 1) the search request $x$, 2) the \textit{top items} of candidate sources $top(\overline{\mathcal{SR}}(x))$, 3) the last primitive action $a_{t-1}$, and 4) the option $o$ that launches the current subtask, and the state vector $\phi(s_t^i)$ is the concatenation of all vectors of each corresponding part.
In order to give an intuitive understanding of \textit{top items}, we take Fig. \ref{slot} as an instance. $top(\overline{\mathcal{SR}}) = [A5; B1; C2]$ before the agent presents the item at 6th slot position, then $top(\overline{\mathcal{SR}}) = [A5; B2; C2]$ after it chooses the source type $B$.
So naturally, the vector of candidate sources $\textbf{e}(top(\overline{\mathcal{SR}}))$ is represented by the concatenation of the top item vector $\textbf{e}(m_j^{top})$ from each selected source.
In addition, similar to \cite{schaul2015universal}, we put options into the low-level state representation as an additional input throughout the low-level RL process, to make the selected sources available for the policy $\pi$.

\subsubsection{Intrinsic reward}

The intrinsic reward $r^i$ is provided for the item presenter to indicate how well a particular subtask is completed. User clicks and transactions are used to form the intrinsic reward, which is computed as below:
\begin{equation}
    r_t^i = \lambda * click + (1 - \lambda) * \min(ln(1 + pay), \delta),
\label{reward:low}
\end{equation}
where $\lambda$ is a weight factor, $click \in \{1, -1\}$ indicates whether the user clicks or skips the current slot, and $pay \in [0, +\infty)$ is the price of the product that the user buys. When the user does not buy anything, the second term is zeroed as $pay=0$. The second term is also clipped by the value $\delta$ to reduce the impact of extreme observations. In addition, if there is no click or transaction during the current subtask, a small negative value is added to the intrinsic reward at the last time step as a penalty.

\subsection{Q Network Framework}

\begin{figure}[!htp]
    \centering
    \includegraphics[width=.85\linewidth]{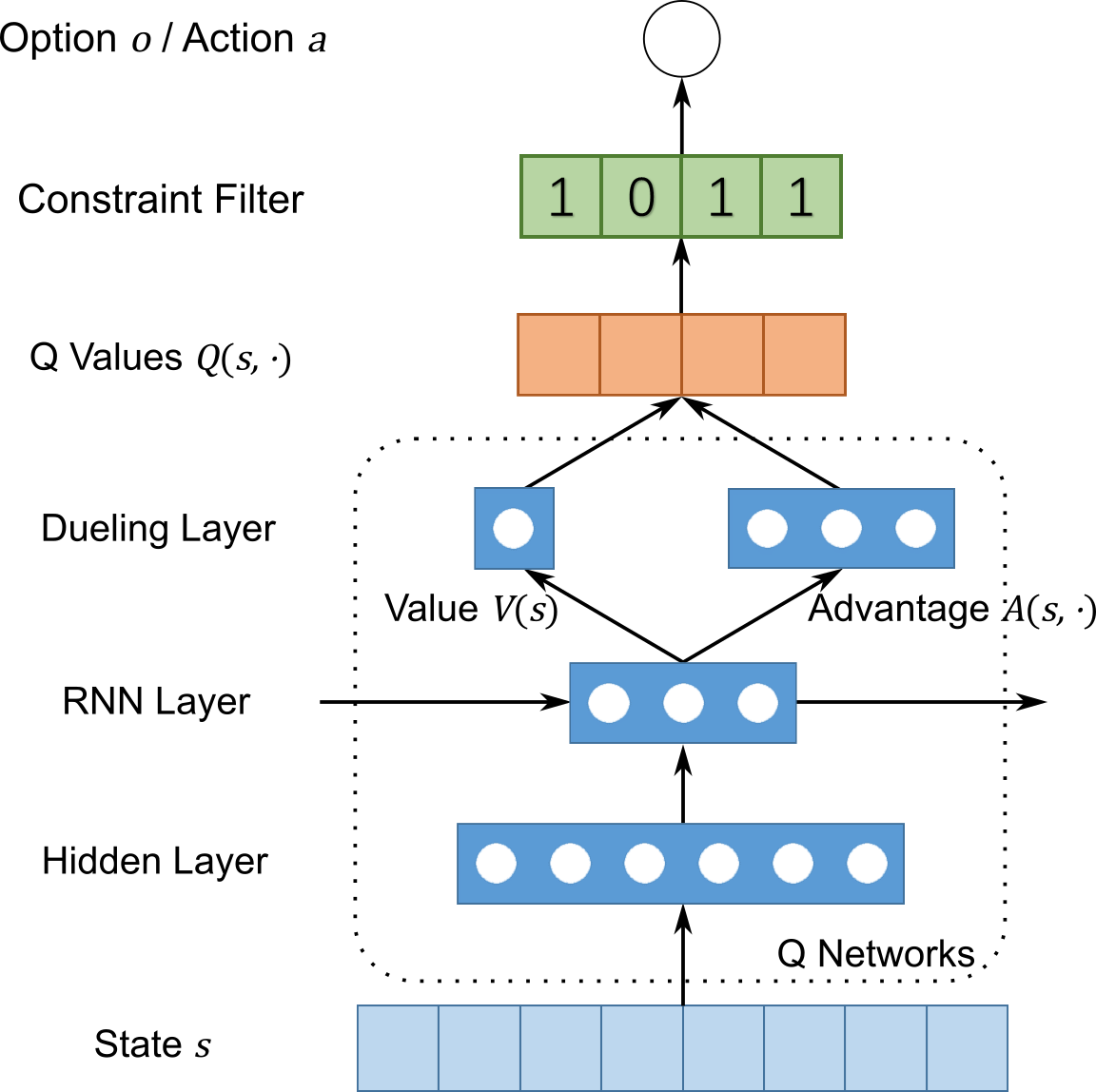}
    \caption{Q network framework used in both high and low level policies.}
    \label{policy}
\end{figure}

Deep Q-learning uses a multi-layer neural networks with parameters $\theta$ to estimate the Q function. Given an $n$-dimensional state $s$, it outputs a vector of Q-values over the actions $Q(s, \cdot; \theta) : \mathbb{R}^n \rightarrow \mathbb{R}^{|\mathcal{A}|}$.

In our algorithm, both high-level and low-level policies leverage the same Q network framework, as shown in Fig. \ref{policy}. We apply DRQN \cite{hausknecht2015deep} by considering the previous context with a recurrent structure to integrate the underlying correlation between the previous state and current observation. Eq. \ref{q:network} refers to the hidden layer and RNN layer in Fig. \ref{policy}.
\begin{equation}
\begin{split}
    \mathbf{w}_t &= \beta(\mathbf{W}_\phi \cdot \phi(s_t) + b_\phi), \\
    \mathbf{h}_t &= GRU(\mathbf{w}_t, \mathbf{h}_{t-1}),
\end{split}
\label{q:network}
\end{equation}
where $\phi(s)$ is the vector representation of the state $s$, $\beta$ is \textit{Leaky ReLU} activation function.

Besides, the dueling architecture \cite{wang2016dueling} is introduced as well to get a more precise policy evaluation in the presence of many similar-valued actions.
It consists of two streams that represent the value and advantage functions, and the two streams are combined afterward to provide a better estimate of Q-values, i.e. the dueling layer of Fig. \ref{policy}.
\begin{equation}
    Q(s, a; \theta) = V(s; \theta) + (A(s, a; \theta) - \frac{1}{|\mathcal{A}|} \sum_{a'} A(s, a'; \theta)).
\label{dueling}
\end{equation}

The \textit{constraint filter} is to explicitly perform an element-wise mask operation on the Q-values to meet some complicated commercial needs.
In this way, commercial constraints in a special situation can be easily satisfied in this framework. For instance, not displaying any blog posts in the first page can be dealt with by the source selector via excluding the corresponding source type from the action space. Similarly, displaying a particular source item at a pre-specified position can be handled by the item presenter via excluding the corresponding slot from the slot sequence.

\subsection{Hierarchical Policy Learning}\label{subsec:optimize}

\begin{algorithm}[!htb]
\DontPrintSemicolon
\KwIn{Item sets with multiple source types $\mathcal{SR}$ and their search services.}
Initialize the two policies with parameters \{ $\theta^e$, $\theta^i$ \} \;
Initialize replay memories \{ $D^e$, $D^i$ \} respectively \;
$\theta^{e-} = \theta^e$, $\theta^{i-} = \theta^i$ \;
\ForEach{user search session}{
    Initialize high-level state $s^e$ with current search request $x$\;
    \Repeat{the session ends}{ \tcp{Page-level source selection}
        $   \text{Set option } o =
            \begin{cases}
                \text{random sample option } o \text{ from } \mathcal{O}, & random() < \epsilon^e \\
                \argmax_{o'} Q^\mu(s^e, o'; \theta^e), & otherwise \\
            \end{cases}
        $

        Initialize low-level state $s^i$ and option duration $l$ according to $o$\;
        \For{$k = 0$ \KwTo $l-1$}{ \tcp{Slot-level item presentation}
            $   \text{Set primitive action } a =
                \begin{cases}
                    \text{random sample action } a \text{ from } \mathcal{A}, & random() < \epsilon^i \\
                    \argmax_{a'} Q^\pi(s^i, o, a'; \theta^i), & otherwise \\
                \end{cases}
            $

            Execute action $a$, observe intrinsic reward $r^i$ (cf. Eq. \ref{reward:low}) and next state ${s'}^i$ \;
            Store transition ($s^i$, $o$, $a$, $r^i$, ${s'}^i$) in $D^i$ \;
            Randomly sample mini-batches from $D^i$ \;
            Perform gradient descent on $L_\pi(\theta^i)$ (cf. Eq. \ref{loss:low} \& \ref{grad:low}) \;
            $s^i = {s'}^i$ \;
            Assign $\theta^{i-} = \theta^i$ every $C^i$ steps \;
        }
        Obtain extrinsic reward $r^e$ (cf. Eq. \ref{reward:high}) and next state ${s'}^e$ \;
        Store transition ($s^e$, $o$, $r^e$, ${s'}^e$) in $D^e$ \;
        Randomly sample mini-batches from $D^e$ \;
        Perform gradient descent on $L_\mu(\theta^e)$ (cf. Eq. \ref{loss:high} \& \ref{grad:high}) \;
        $s^e = {s'}^e$ \;
        Assign $\theta^{e-} = \theta^e$ every $C^e$ steps \;
    }
}
\caption{Deep hierarchical policy learning for aggregated search}
\label{algorithm}
\end{algorithm}

Standard DQN is used to minimize the squared error between the target (Bellman optimality condition) $r + \gamma \max_{a'} Q(s', a'; \theta)$ and its estimate $\hat{Q}(s, a; \theta)$. In the context of HRL, the top-level policy $\mu$ tries to minimize the following loss function at each time step $t$:
\begin{equation}
\begin{split}
    L_\mu(\theta^e_t) &= \mathbb{E}_{(s, o, r, s') \sim D^e} [(y^e_t - Q_\mu(s, o; \theta^e_t))^2], \\
    y^e_t &= r + \gamma^l Q_\mu(s', \argmax_{o'} Q_\mu(s', o'; \theta^e_t) ; \theta^{e-}_t),
\end{split}
\label{loss:high}
\end{equation}
where $l$ is the number of steps the option $o$ lasts. Here we apply experience buffer \cite{mnih2013playing} to smooth over changes in the data distribution. We also employ a target Q network \cite{mnih2015human} with parameters $\theta^-$ to decrease the possibility of divergence or oscillations during the training process, and double Q-learning \cite{hasselt2016deep} to reduce the observed over-estimations caused by traditional Q-learning. Similarly, we learn the low-level policy $\pi$ by minimizing the following loss function:
\begin{equation}
\begin{split}
    L_\pi(\theta^i_t) &= \mathbb{E}_{(s, o, a, r, s') \sim D^i} [(y^i_t - Q_\pi(s, o, a; \theta^i_t))^2], \\
    y^i_t &= r + \gamma Q_\pi(s', o, \argmax_{a'} Q_\pi(s', o, a'; \theta^i_t); \theta^{i-}_t),
\end{split}
\label{loss:low}
\end{equation}
given the option $o$ from $\mu$.

Differentiating the loss function with respect to the weights, we arrive at the following gradient for the high-level policy:
\begin{equation}
    \nabla_{\theta^e_t} L_\mu = \mathbb{E}_{(s, o, r, s') \sim D^e} [(y^e_t - Q_\mu(s, o; \theta^e_t)) \nabla_{\theta^e_t} Q_\mu(s, o; \theta^e_t)],
\label{grad:high}
\end{equation}
and similarly, the gradient for the low-level policy yields as below:
\begin{equation}
    \nabla_{\theta^i_t} L_\pi = \mathbb{E}_{(s, o, a, r, s') \sim D^i} [(y^i_t - Q_\pi(s, o, a; \theta^i_t)) \nabla_{\theta^i_t} Q_\pi(s, o, a; \theta^i_t)].
\label{grad:low}
\end{equation}

Note that the initial hidden state of RNN layer should be carried forward from its previous values $\mathbf{h}_{t-1}$ when sampling actions, but we zero it at the start of the update so that we do not need to save $\mathbf{h}_{t-1}$ into replay buffers. The convergence and performance of such complexity reduction are guaranteed \cite{hausknecht2015deep}. In addition, we clip the error term through \textit{Huber Loss} to improve the stability of the algorithm.

The entire training process is described in Algo. \ref{algorithm}.

\section{Experimental Setup}\label{sec:experiment_setup}

To evaluate the performance of our proposed approach, we carried out experiments on the Taobao Search platform, which is one of the largest E-commerce search services in the world, with over 1 billion user clicks every day.
All models in this paper adopt the online-learning paradigm and are trained on the hourly real-time search log streaming data.

The online experiment methodology we adopted is called bucket testing, also known as A/B test.
In the bucket testing system of Taobao Search, several test buckets are set up.
And all users are randomly hashed into these buckets based on their user ids.
Each bucket has the same number of users, and the same distribution of users.
Then each algorithm is deployed on one bucket.
The performance of an algorithm is estimated using the metrics calculated on the bucket it is deployed on.
In this paper, our online bucket testing lasted for two weeks, a period long enough to ensure the statistical stability of our test results.

\subsection{Features and Parameter Settings}

We aggregated two verticals, namely topic and group verticals, into the results of product search.
For each RL policy in our model, the components of state $s$ (i.e. search request $x$, search source results $\mathcal{SR}(x)$, etc.) are represented by a number of features, which mainly contains 1) the query features: including the word-segmented tokens of the query; 2) user features: including user gender, age, items user clicked in the previous page; 3) page features: the current page number; 4) source features obtained from each search services, including the type and title of a topic group or a blog post.

The parameter settings for the high-level source selector and low-level item presenter are provided in Table \ref{hyper}.

\begin{table}[!htb]
    \centering
    \begin{tabular}{ccc}
    \toprule
    \multirow{2}{*}[-0.04in]{Hyper-parameter}  & \multicolumn{2}{c}{Setting} \\
    \cmidrule(lr){2-3}
    & Source selector & Item presenter \\
    \midrule
    Learning rate     & 1e-2 & 1e-4  \\
    Optimization algorithm & RMSProp & RMSProp \\
    Memory $D$ size & 5e4 & 5e5 \\
    Mini-batch size & 32 & 32 \\
    Target $\theta^-$ update period $C$ & 1e3 & 1e4 \\
    State feature vector size & 48 & 56 \\
    Hidden layer size & 28 & 24 \\
    RNN layer size & 16 & 12 \\
    Discount factor $\gamma$ & \multicolumn{2}{c}{0.95} \\
    Weight factor $\lambda$ in Eq. \ref{reward:low} & - & 0.3 \\
    Clipped to $\delta$ in Eq. \ref{reward:low} & - & 3 \\
    \bottomrule
    \end{tabular}
    \caption{Hyper-parameter settings}
    \label{hyper}
\end{table}

\begin{table*}[!htp]
    \centering
    \begin{tabular}{lcccccccccccc}
        \toprule
        \multirow{3}{*}[-0.04in]{Method} &\multicolumn{6}{c}{Topic group} &\multicolumn{6}{c}{Blog post} \\
        \cmidrule(lr){2-7} \cmidrule(lr){8-13}
        &\multicolumn{2}{c}{CTR} & \multicolumn{2}{c}{ADT} & \multicolumn{2}{c}{COV} &\multicolumn{2}{c}{CTR} & \multicolumn{2}{c}{ADT} & \multicolumn{2}{c}{COV} \\
        \cmidrule(lr){2-3} \cmidrule(lr){4-5} \cmidrule(lr){6-7} \cmidrule(lr){8-9} \cmidrule(lr){10-11} \cmidrule(lr){12-13}
            & value \small{(1e-2)}    & gain    & value    & gain & value \small{(1e-2)}    & gain & value \small{(1e-2)}    & gain & value    & gain & value \small{(1e-2)}    & gain \\
        \midrule
        Rule & 5.24 & - & 10.56 & - & 5.60 & - & 3.43 & - & 75.78 & - & 6.47 & - \\
        Flat RL &  6.08 & 15.99\% & 10.62 & 0.54\% & 5.98 & 6.79\% & 3.92 & 14.29\% & 76.69 & 1.20\% & 6.76 & 4.48\% \\
        BC+RM & 5.59 & 6.76\% & 10.68 & 1.09\% & 6.03 & 7.68\% & 3.69 & 7.58\% & 76.99 & 1.59\% & 6.75 & 4.33\% \\
        BC+RL & 5.60 & 6.77\% & 10.66 & 0.92\% & 6.00 & 7.32\% & 3.73 & 8.75\% & 76.69 & 1.21\% & 6.82 & 5.41\% \\
        RL+RM & 5.52 & 5.29\% & 10.75 & 1.75\% & 6.09 & 8.75\% & 3.60 & 6.12\% & 76.46 & 0.91\% & 6.87 & 6.18\% \\
        HRL & 7.34 & \textbf{40.07}\% & 10.65 & 0.89\% & 5.95 & 6.25\% & 4.25 & \textbf{23.91}\% & 76.53 & 0.98\% & 6.86 & 6.03\% \\
        \bottomrule
    \end{tabular}
    \caption{Metrics of online bucket testing on two verticals.}
    \label{tab:main}
\end{table*}

\subsection{Baselines and Evaluation Metrics}\label{subsec:baselines}

In order to show the effectiveness of our HRL model, we implemented several aggregation methods for comparison, which can be grouped into two categories: \textit{rule-based} and \textit{learning to aggregate} methods, while the latter can be separated into two types: \textit{one-stage} and \textit{two-stage} methods according to the process flow. The \textit{one-stage} method is introduced to verify the hierarchical decomposition in the aggregate search, and the \textit{two-stage} method is carried out to validate the RL formulation in each subtask.
The details of our baselines are introduced as follows:

\textit{Rule}: a rule-based method, where the rule is manually made for search aggregation according to abundant experience in search product design. The rule claims that the first page should not present aggregated results, and the topic and blog verticals will be alternately displayed at fixed positions from the second page when a topic group or a blog post is available from backends. In a page, the rule displays a topic group at the 4th position, and a blog post at the 9th position. Note that this simple rule is the benchmark for comparison with other methods.

\begin{figure}[!htb]
\centering
\includegraphics[width=\linewidth]{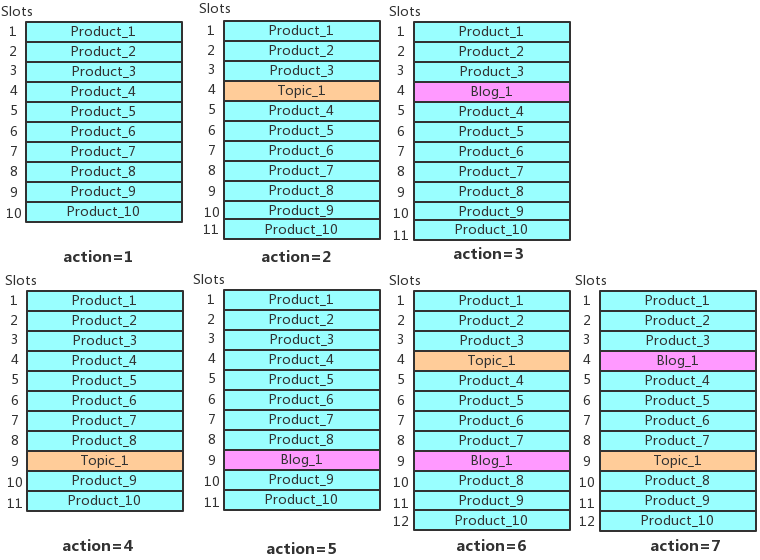}
\caption{The SERP templates for our 1-stage DQN baseline.}
\label{Fig:template}
\end{figure}

\textit{Flat RL}: a one-stage method. In this paradigm, the aggregation task is simplified into one stage: for each aggregation request, just choose the optimal template from a set of predefined templates. Seven templates are used in this method and are provided in Fig. \ref{Fig:template}. Each template clarifies exactly what to fill in each slot. These templates are designed by human experts familiar with Taobao Search to ensure good customer experience. The aggregation problem is solved by a single DQN agent. The state representation and the reward of this DQN are the same with the high-level state and the extrinsic reward defined in subsection \ref{subsec:high} respectively.

\textit{BC+RM}: a two-stage method similar to \cite{ponnuswami2011composition}.
For \textit{vertical selection}, it uses a binary classifier for each source to decide whether to present this source in a page. The binary classifiers are Neural Networks with 3 hidden layers.
The training data for the binary classifiers come from online search log, where each page view is a sample. For a vertical in the page, if it is clicked by a user, then it is a positive sample, otherwise a negative sample.
For \textit{vertical presentation}, it uses a 3-hidden-layer Neural Network regression model to score each item. The regression target of the regression model is the same with the intrinsic reward in Eq. \ref{reward:low}.

\textit{BC+RL}: a two-stage method similar to \textit{HRL}, which replaces the high-level source selector with multiple binary classifiers as described in \textit{BC+RM}, but the RL policy for the item presenter remains unchanged.

\textit{RL+RM}: a two-stage method similar to \textit{HRL}, which replaces the low-level item presenter with a regression model as described in \textit{BC+RM}, but the RL policy for the source selector remains unchanged.

Note that all \textit{learning to aggregate} methods were pre-trained on offline data by \textit{behavioral cloning} prior to online experiments, so as to mitigate the slow learning and poor performance in the early stage of online training. The offline data, which contain 60,000 search sessions covering a broad query topics with an average of 11.3 items per page and 13.4 pages per session, were collected from the user search logs on the Taobao platform.

To evaluate the performance of the algorithms, we use some common evaluation metrics in search aggregation, including:

\textit{Click Through Rate} (CTR): the ratio of users who click on a specific vertical to the number of total users who view the vertical on a page.

\textit{Average Dwell Time} (ADT): the mean value of a user's dwell time in seconds on a specific vertical.

\textit{Coverage} (COV): the ratio of the number of slots occupied by a specific vertical to the total number of slots.

Moreover, we also include an important metric in the E-commerce industry: the \textit{Gross Merchandise Volume} (GMV) of each search service, which is the total revenue of sales made by the product search, or induced by a vertical in SERPs within a fixed time period.

\section{Results}

\subsection{Online Performance}
We show the results of online bucket testing in terms of CTR, ADT and COV of different verticals in Table \ref{tab:main}.
Among these three metrics, the most important one is CTR, which indicates whether users clicked the verticals in the aggregated SERP.
For each metric in Table \ref{tab:main}, we present its value in the "value" column, and its relative improvement over the \textit{Rule} baseline in the "gain" column.
The results in Table \ref{tab:main} show that all the baselines and our HRL model achieve similar improvement rate on ADT and COV.
However, it is worth noting that the result on CTR differs remarkably among all the methods.

In general, \textbf{HRL} achieves much better CTR improvement than all the baselines on both topic and blog verticals.
The comparison to \textbf{Flat RL} shows that our hierarchical decomposition of the RL process is very effective, achieving better CTR from the historical user behaviors across different pages as well as the immediate user feedback in the current page.
And the result also indicates that it is more sensible to allow a model to choose from any positions when displaying heterogeneous items, rather than restrict the positions where a vertical can be presented in a page.
While the \textit{one-stage} method can only use limited aggregation templates to meet the trade-off between the model performance and the exponential number growth of templates due to the number of sources, the hierarchical decomposition of \textit{HRL} overcomes the exponentially large action space problem and allows the agent to explore all possible source combination and item permutation.

The comparisons to \textit{BC+RM}, \textit{BC+RL}, \textit{RL+RM} show that the sequential RL modeling in both high-level and low-level aggregation subtasks is critical for a good performance. A common feature of these discriminant models is that they all follow the vertical selection and presentation pipeline, but make point-wise decisions, without the passing of states between sequential decisions as in \textit{HRL}.
\textbf{BC+RL} considers the user feedback only based on the current page, while \textit{HRL} tracks the user behaviors from page to page to understand user intents. The result verifies that it is essential to capture sequential patterns of user behavior in aggregated E-commerce search.
Similarly, the comparison to \textbf{RL+RM} demonstrates that \textit{HRL} also consistently outperforms the baseline that simply ranks items from heterogeneous sources since it suffers from relevance ranking issue by giving scores to each vertical, which concludes that our slot filling process aggregates a more reasonable SERP.

Another important metric, the GMV of the aggregated E-commerce search, that we care about is shown in Table \ref{tab:GMV_main}.
It should be noted that, due to Alibaba’s business policy, we temporarily cannot expose the absolute values of GMV. Hence, we report relative GMV increase over \textit{Rule} instead, and this will not affect performance comparison.
Because the GMV of aggregated search is the total merchandised volume of all sources, we report not only the GMV of topic and blog verticals, but also the GMV of product search in Table \ref{tab:GMV_main}.

\begin{table}[!htb]
    \centering
        \begin{tabular}{lccc}
        \toprule
        Method & Topic group & Blog post & Products\\
        \midrule
        Flat RL & 38.4\% & 12.31\% & -0.47\% \\
        BC+RM & 19.32\% & 5.83\% & -0.41\% \\
        BC+RL & 20.78\% & 6.00\% & -0.37\% \\
        RL+RM & 17.60\% & 5.54\% & -0.20\% \\
        HRL & \textbf{56.49}\% & \textbf{16.93}\%  & \textbf{0.56}\% \\
        \bottomrule
        \end{tabular}
    \caption{The relative GMV increase over \textit{Rule}. The GMV of a vertical is computed through the sales value guided by the presented items of that vertical.}
    \label{tab:GMV_main}
\end{table}

As shown in Table \ref{tab:GMV_main}, \textit{HRL} is the only method that increases the GMV of product search.
Although other baselines achieve GMV increase on the topic and blog verticals, their GMVs on product search all decrease.
For example, \textit{Flat RL} brings a marked GMV improvement on topic and blog verticals, at the expense of the largest GMV drop on product search though.
This means that \textit{HRL} provides the best user experience in its aggregated SERP, because even a growth on the COV of topic and blog verticals does not reduce users' purchases in product search.
Moreover, \textit{HRL} also achieves the best GMV increase on the topic and blog verticals, leading to the best overall GMV increase among all methods.

\begin{figure}[!htb]
    \centering
    \includegraphics[width=\linewidth]{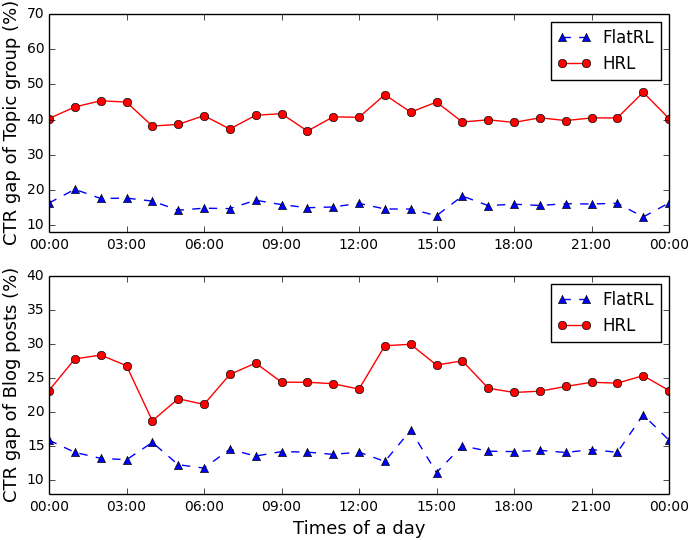}
    \caption{The relative CTR increase over \textit{Rule} in different hours. Each mark indicates the result in the time span between the current time and next full hour.}
    \label{relative}
\end{figure}

To provide more insights on the online experiments, we group the two-week statistics by different hours of a day to observe the performance variation over time. For simplicity, we only compare the relative CTR increase of the two most competitive methods: \textit{Flat RL} and \textit{HRL}. Fig. \ref{relative} shows the trend of their CTR increase over the baseline \textit{Rule} during different time periods of a day. It can be seen that the CTR increase does not variate too much at different hours of a day, and \textit{HRL} has consistently better CTR improvement than \textit{Flat RL}.

\subsection{User Evaluation}
In order to study the user experience change caused by our HRL aggregation method, we invite 50 users of Taobao Search to evaluate the quality of our aggregated result against that of the rule-based baseline: \textit{Rule}, which is the old online aggregation method of Taobao Search and has been online serving for several years.
The evaluation is performed on a fixed query set: 200 popular queries covering 13 salient product categories in Taobao.
This query set is selected by third-party experts on Taobao platform.
A user is asked to search each query in both our and the old aggregation environments simultaneously, without knowing which environment is the old one.
And the aggregated results of the two environments, designated by A and B, are presented to the user side-by-side.
The user can browse the results, click items, scroll to the next page, and so on.
And the user provides a grade from \{"System A better", "Same/cannot judge", "System B better"\} on each vertical for this query.
Then we translate the user's grade to \{"New System better", "Same/cannot judge", "Old System better"\}, where the "New System" refers to \textit{HRL} and the "Old System" refers to \textit{Rule}.
The distribution of user grades on each vertical is shown in Table \ref{tab:user_eval}.

\begin{table}[!htb]
    \centering
        \begin{tabular}{lccc}
        \toprule
        Vertical & New system better & Same & Old system better \\
        \midrule
        Topic & 30.35\% & 67.98\% & 1.67\% \\
        Blog & 26.14\% & 72.53\% & 1.33\% \\
        \bottomrule
        \end{tabular}
    \caption{The distributions of user evaluation grades.\label{tab:user_eval}}
\end{table}

The results in Table \ref{tab:user_eval} show that user experience grades are the same on about 70\% of the queries.
In spite of this, \textit{HRL} provides better user experience than the baseline most of the times on the rest 30\% of the queries.

To understand how \textit{HRL} improves user experience, every time when a user thinks the new system is better, we inquire about his reason for this preference.
And we also ask them to group their reasons into three groups as follows:
\begin{itemize}
	\item Better \textbf{behavior relevance}: In \textit{HRL}, the presented verticals are more relevant to the user's previous behaviors.
	\item Better \textbf{query relevance}: In \textit{HRL}, the presented verticals are more relevant to the query.
	\item \textbf{Others}: other reasons.
\end{itemize}

\begin{table}[!htb]
    \centering
        \begin{tabular}{lccc}
        \toprule
        Vertical & Behavior relevance & Query relevance & Others \\
        \midrule
        Topic & 42\% & 33\% & 25\% \\
        Blog & 37\% & 35\% & 28\% \\
        \bottomrule
        \end{tabular}
    \caption{Distributions of reasons why new system is better.\label{tab:user_eval_reason}}
\end{table}

The results in Table \ref{tab:user_eval_reason} show that \textit{HRL} contributes to both user behavior relevance and query relevance. We present two cases here to give some clue that how user experience is improved.

A case for better user behavior relevance is as follows: a user issues the query "running shoes", and clicks several products with the same brand: "Nike" on the first page, then on the next page the topic group titled "Nike shops with discounts" is presented at the second position in our HRL aggregated result, which is very relevant to the user's behavior.
In contrast, no topic group is presented by the baseline in the aggregated result on the next page.
In \textit{HRL}, user behaviors in previous pages are encoded into the RNN layer hidden states in the Q network shown in Fig. \ref{policy}.
So user behaviors in previous pages are passed on to the next page for decision making in \textit{HRL}, which helps perceive user behavior relevance.
Besides, \textit{HRL} can also capture sequential patterns of user behavior by learning from the long-delayed user feedback across the whole user search session.

A case for better query relevance is as follows: a user issues the query "women's long dress autumn Vero Moda", which is very specific in brand, season and style.
The blog post titled "Introduction to the world of women's dresses", which gives a general introduction to women's dresses, is not presented in the aggregated result of \textit{HRL}, whereas it is presented in the baseline's aggregated result.
The user prefers our aggregated result because she thinks the blog post is too general thus not meeting her specific needs.
With query and blog title features as input, \textit{HRL} is able to identify that the general introduction of the blog post is not relevant enough to the query she inputs.

\subsection{Training Analysis}

To offer a detail training analysis of HRL, we investigate the training procedure of the high-level RL process that takes multi-step actions. Two strategies are designed for the source selector. (\RNum{1}) One is the proposed one in section \ref{subsec:high} \& \ref{subsec:optimize} that the extrinsic reward is the mean value of the intrinsic return from the item presenter, i.e. the same as Eq. \ref{reward:high}, and the optimization target of Q network ($y_t^e$ in Eq. \ref{loss:high}) is discounted by $\gamma^l$. (\RNum{2}) The other is to regard the high-level RL process as a simple RL process that takes one-step actions. Under this setting, the extrinsic reward is directly equivalent to the mean value of the intrinsic rewards, i.e. $r_t^e = \frac{1}{l} \sum_{k=0}^{l-1} r_{t+k}^i$, then the training target of source selector policy in Eq. \ref{loss:high} is similar to that in Eq. \ref{loss:low}: $y^e_t = r + \gamma Q_\mu(s', \argmax_{o'} Q_\mu(s', o'; \theta^e_t) ; \theta^{e-}_t)$.

\begin{figure}[!htb]
    \centering
    \includegraphics[width=\linewidth]{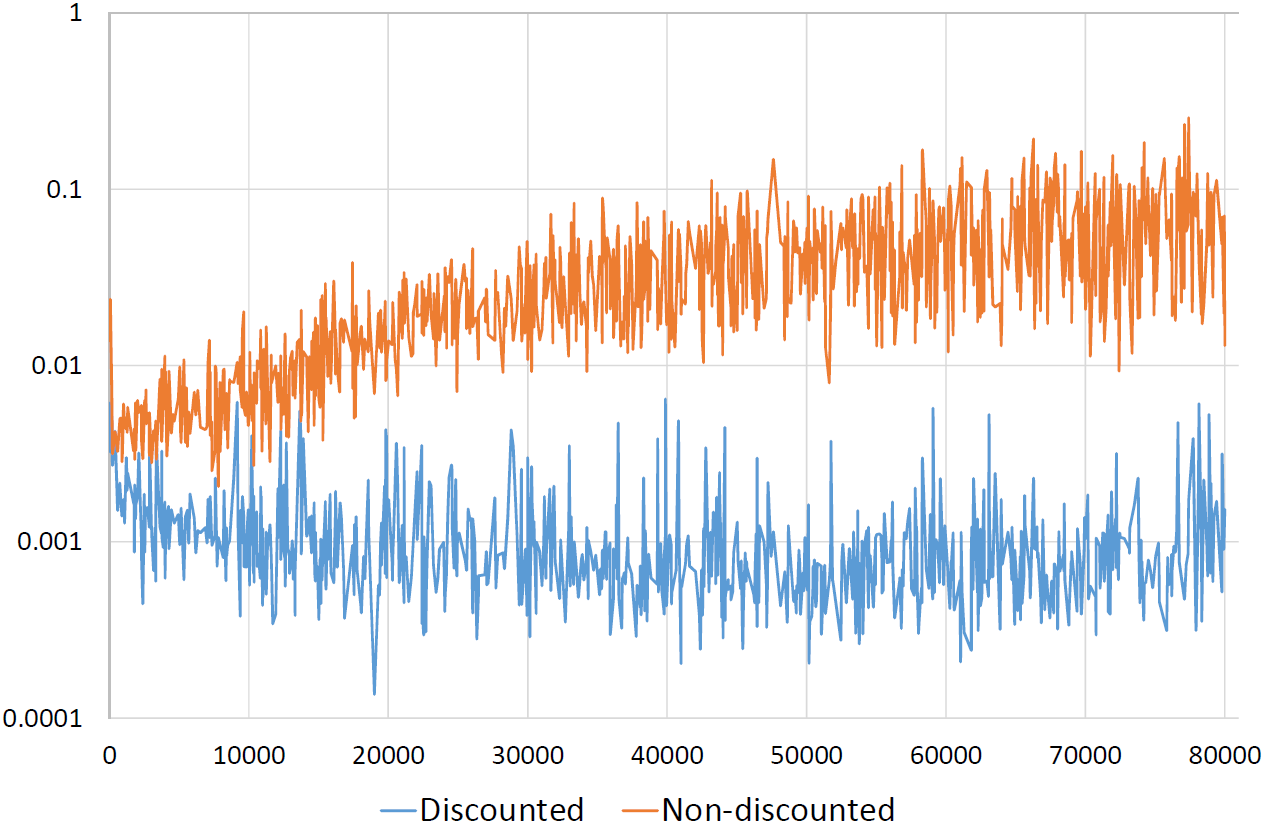}
    \caption{Loss curve of high-level source selector between different training strategies. Blue curve for strategy (\RNum{1}) and orange curve for strategy (\RNum{2}). The horizontal axis is the training iteration, and the vertical axis is the loss value in log scale.}
    \label{fig:high_loss}
\end{figure}

The learning curves of the two strategies are demonstrated in Fig. \ref{fig:high_loss}, where the blue and orange curves correspond to strategy (\RNum{1}) and (\RNum{2}) respectively. It is clear that the proposed strategy (\RNum{1}) converges well as the training procedure proceeds while the strategy (\RNum{2}) diverges dramatically. This result shows that it is necessary to handle a multi-step RL process in a proper way to guarantee its convergence. From another perspective, the strategy (\RNum{1}) puts more weight on the items presented in the top slots, which is consistent with the position bias in most search area that top-ranked search results attract more clicks, so the agent can better understand the sequential patterns of user behavior and learn from such data effectively.

\section{Conclusion and Discussion}

We have studied the search aggregation problem in a large E-commerce search service and proposed an effective HRL aggregation method.
Different from web search aggregation, the E-commerce aggregation needs to be performed for each page, leading to multiple and sequential aggregation decisions for one query.
In light of this, we propose a hierarchical aggregation model for E-commerce search aggregation.
Our HRL model naturally fits the two-stage decomposition of the aggregated search task: a high-level RL for source selection and a low-level RL for item presentation.
The source selector models user interactions on aggregated results across pages to capture the sequential patterns of user behavior, and the item presenter formulates a slot filling process that sequentially presents the heterogeneous items to avoid the relevance ranking issue.
In this manner, the HRL model can fully learn from user feedback in the current page as well as in the entire search session.

We perform online bucket testing on Taobao Search platform and compare our HRL method with several baselines.
The results show that our method contributes a substantial improvement over the baselines for all verticals.
Moreover, it also boosts the GMV of product search and achieves the best overall GMV among all the models.
User study also demonstrates that it improves user experience by displaying better aggregated results that are relevant to user behaviors and needs.
Note that our method can easily combine new verticals into an aggregated search system.

One limitation of the proposed algorithm may lie in the automatic offline evaluation. All we can get from the offline data is the records of user behaviors on the presented results, so it is difficult to infer the user feedback on the unseen SERPs without full annotations of the heterogeneous item relevance as we train our model in a RL setting. However, sufficient experiments including bucket testing and user study indicate that our method indeed shows its effectiveness on the search aggregation.
Future research will be focused on the application of our HRL model in different E-commerce verticals and other scenarios.

\section*{Acknowledgements}
This work was jointly supported by the National Key R\&D Program of China (Grant No. 2018YFC0830200) and the National Science Foundation of China (Grant No.61876096/61332007). We would also like to thank Mr. Xiaoyi Zeng, our colleague in Alibaba, for constant support and encouragement.

\bibliographystyle{ACM-Reference-Format}
\bibliography{www}

\end{document}